# Technological implementation of a photonic Bier-Glass cavity

Jonathan Jurkat[1,*], Magdalena Moczała-Dusanowska[1], Martin Arentoft Jacobsen[2], Ana Predojević[3], Tobias Huber[1], Niels Gregersen[2], Sven Höfling[1], and Christian Schneider[1,4]

1 - Technische Physik and Wilhelm Conrad Röntgen Research Center for Complex Material Systems, Physikalisches Institut, Würzburg University, Am Hubland, Würzburg, Germany

2 - DTU Fotonik, Department of Photonics Engineering, Technical University of Denmark, Ørsteds Plads 343, 2800 Kongens Lyngby, Denmark

3 - Department of Physics, Stockholm University, 106 91 Stockholm, Sweden

4 – Institute of Physics, University of Oldenburg, 26129 Oldenburg, Germany

* Corresponding author: jonathan.jurkat@physik.uni-wuerzburg.de

In this paper, we introduce a novel quantum photonic device, which we term photonic Bier-Glass cavity. We discuss its fabrication and functionality, which is based on the coupling of integrated In(Ga)As quantum dots to a broadband photonic cavity resonance. By design, the device architecture uniquely combines the Purcell enhancement of a photonic micropillar structure with broadband photonic mode shaping of a vertical, tapered waveguide, making it an interesting candidate to enable the efficient extraction of entangled photon pairs. We detail the epitaxial growth of the heterostructure and the necessary lithography steps to approach a GaAs-based photonic device with a height exceeding 15 µm, supported on a pedestal that can be as thin as 20 nm. We present an optical characterization, which confirms the presence of broadband optical resonances, in conjunction with amplified spontaneous emission of single photons.

**Introduction**

The preparation of single photons and entangled photon pairs (EPPs) is a critical resource in the fields of quantum optics, quantum metrology and quantum information [1–3] . Quantum dots (QDs) embedded in microcavities are a promising candidate to create such non-classical light states. The spontaneous emission enhancement experienced by the QD in a cavity is a valuable tool to reach very high photon coupling efficiencies into resonating modes and to boost the overall device efficiencies (the photon extraction and collection efficiency) beyond 75%[4–6]. Furthermore, the spontaneous emission enhancement is key to mitigate the effects of pure dephasing on the quantum emitter by controlling the radiative transition lifetime. This enables the generation of highly indistinguishable photons, without the need for strong spectral filtering, which would decrease the system efficiency[7–9]. However, since in most implementations of coupled QD-cavity systems, the mode volumes are in the order of $\left(\lambda/n\right)^3$, relatively high Q factors ($10^3$ or higher) are needed to facilitate a notable spontaneous emission enhancement. However, under those conditions, the Purcell effect becomes prominent within a small bandwidth only, which is prohibitive for the efficient extraction of entangled photon pairs. Broadband approaches based on photonic waveguides have been introduced to implement efficient single photon sources [10] and photon pair sources with improved characteristics [11,12] based on III/V quantum dots. However, thus far, it turned out utmost challenging, both from the modelling as well as the technology development, to combine the broadband performance of a photonic waveguide with the spontaneous emission enhancement of a microcavity.

Here, we address this problem, following a modification of a device suggestion proposed by Gregersen et al. [13]. By integrating a distributed Bragg-reflector (DBR) in a GaAs-based photonic trumpet, it was suggested to combine a modest quality factor cavity, supporting Purcell factors of up to 3, with the photonic waveguide effect that yields suppressed emission into leaky modes. While the initial suggestions considered a metallic back-mirror [13], we implement a second DBR to mimic the symmetry of a DBR based micropillar. This also brings the advantage of a fully epitaxial structure, without the necessity of complicated wafer-bonding steps.

Our re-designed device resembles the shape of a german Bier-Glass. This shape consists of a taper section which contains the waveguide and the DBRs and a foot that shows an inverted taper. A device of this shape theoretically supports an extraction efficiency up to 0.725 together with a Purcell factor up to 3. We demonstrate the modeling as well as the necessary technology for fabrication of these seemingly fragile object and our optical characterization verifies the presence of optical resonances, as well as pronounced, bright QD emission signals.

**Modeling**

We first perform a numerical investigation of the performance of the Bier-Glass geometry. A sketch of the simulated device is depicted in Figure 1a). For a wavelength of λ= 925 nm , the cavity (DBR layer) optical thickness is chosen as $\lambda/n_{eff}$ ($\lambda/4n_{eff}$) taking into account the diameter-dependence of the effective refractive index $n_{eff}$ [13–15]. To ensure optimal transmission to a Gaussian profile of a high numerical aperture (NA) lens (0.8 NA), the top DBR is followed by a taper and an anti-reflective (AR)coating, where $h_{taper}$ is chosen as the smallest value ensuring that $d_{top} \geq 2$ μm.

The simulations were performed using a Fourier modal method[16] with a true open geometry boundary condition [17] combined with a standard scattering matrix formalism [18]. In Figure 1b) we plot the Q-factor off the cavity as a function of its diameter. For diameters below 2 μm we observe fast oscillations before flattening to a constant value, which is a similar behavior seen in vertical micropillars [14,19,20]. Then Figure 1c) shows the Purcell factor which reaches a value of approximately 3 for a diameter of around 0.55 μm. As $F_p \propto \frac{Q}{V}$, where V is the mode volume, the Purcell factor will decrease as we increase the diameter due to the increased mode volume. The $\beta$-factor seen in Figure 1d), which describes the emission fraction into the cavity mode, follows the tendency of the Purcell factor and reaches a maximum value of $\beta$ = 0.854 at $d_{cav}$ = 700 nm. Finally, we have the total coupling efficiency to a lens, which follows the behavior of the $\beta$-factor, and reaches a maximum value of $\varepsilon$ = 0.725 (Figure 1e)) at $d_{cav}$ = 700 nm.

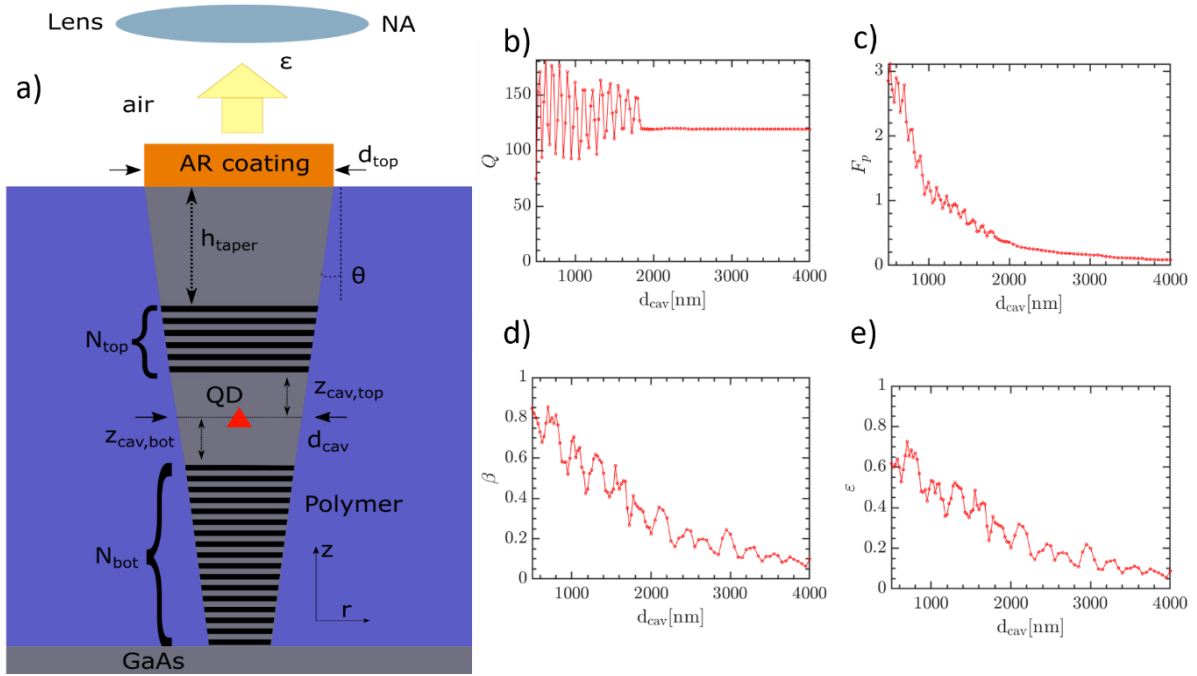

Figure 1 a) Sketch of the simulated device. A quantum dot (QD) is centered in a λ-cavity ($\lambda = 925 nm$) surrounded by DBR mirrors with $N_{bot}$=16 ($N_{top}$=8) layer pairs in the bottom (top) mirror. The materials used are GaAs/Al$_{0.85}$Ga$_{0.15}$As. The structure has a constant sidewall angle of $\theta = 3°$. Above the top DBR, the structure features a taper of height, $h_{taper}$, and an AR-coating. The refractive indices are: $n_{GaAs} = 3.48$ [21], $n_{AlGaAs} = 2.99$ [21], $n_{AR} = 1.99$ [22] and $n_{clad} = 1.57$ [23]. b) Q-factor, c) Purcell factor, d) β-factor and e) Efficiency, ε, as a function $d_{cav}$.

**Device growth**

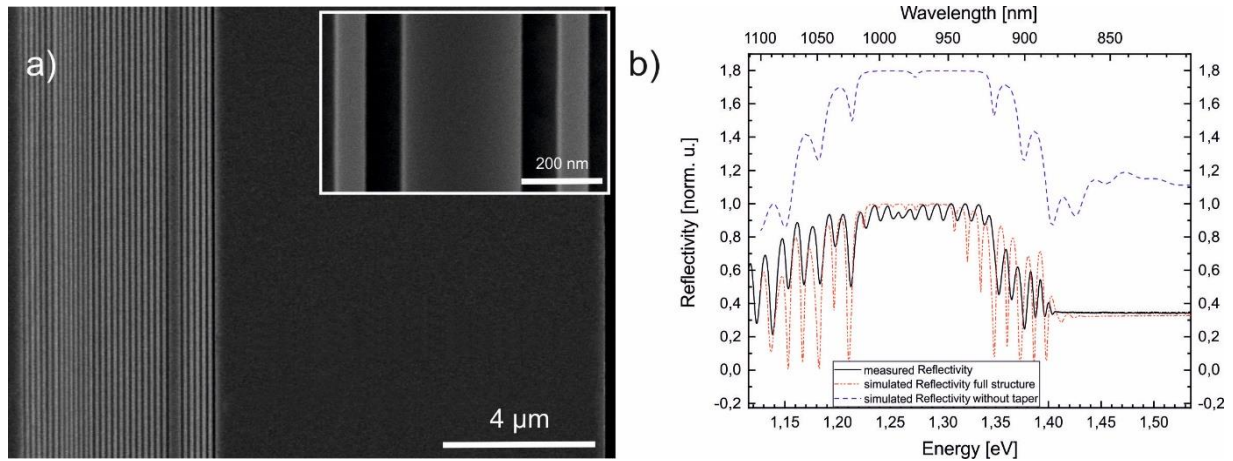

Figure 2) a) Cleaved edge SEM Picture of the planar wafer. Two DBR sections (26 and 6 mirror pairs) separated by the cavity layer, which contains the InAs QDs, followed by 10 µm GaAs (left to right) The cavity layer and the surrounding mirrorpairs can be seen in the islet. b) Measured reflectivity of the planar structure (black), compared to the simulated reflectivity of the full structure (red) and the simulated structure without the 10 µm GaAs top layer (blue with an offset of 0.8) .

The epitaxial structure for the devices was grown via molecular beam epitaxy (MBE) on a 001 oriented GaAs wafer. The layer sequence has been optimized for a Bier-Glass shape broadband photonic cavity. First, a 300 nm GaAs buffer was grown to smoothen the surface followed by 26 bottom DBR pairs of $Al_{0.85}Ga_{0.15}As$ and by a 287 nm thick GaAs cavity layer. The buffer layer is not shown in this scanning electron microscope (SEM) image. A detail of the cavity layer with the previous and following mirror pair is shown in the inset of Figure 1a). The cavity layer contains In(Ga)As QDs in the middle where the electric field forms an antinode. The second DBR consists of 6 mirror pairs. The structure is finished by a GaAs layer that is nominally 10 µm thick. The layer structure was designed to develop a stopband between 920 nm and 1020 nm. The red curve in Figure 2b) depicts a transfer matrix simulation of the full structure, while the black curve is the experimentally measured reflectivity spectrum. The oscillations in the stopband are caused by the 10 µm GaAs layer, which is forming an additional Fabry-Pérot cavity, with the top DBR and the surface to air as its mirrors. A simulation of the structure without the thick GaAs layer (blue curve in Figure 2b), shows the reflectivity of the DBR structure with the characteristic cavity resonance at 973 nm.

**Device fabrication**

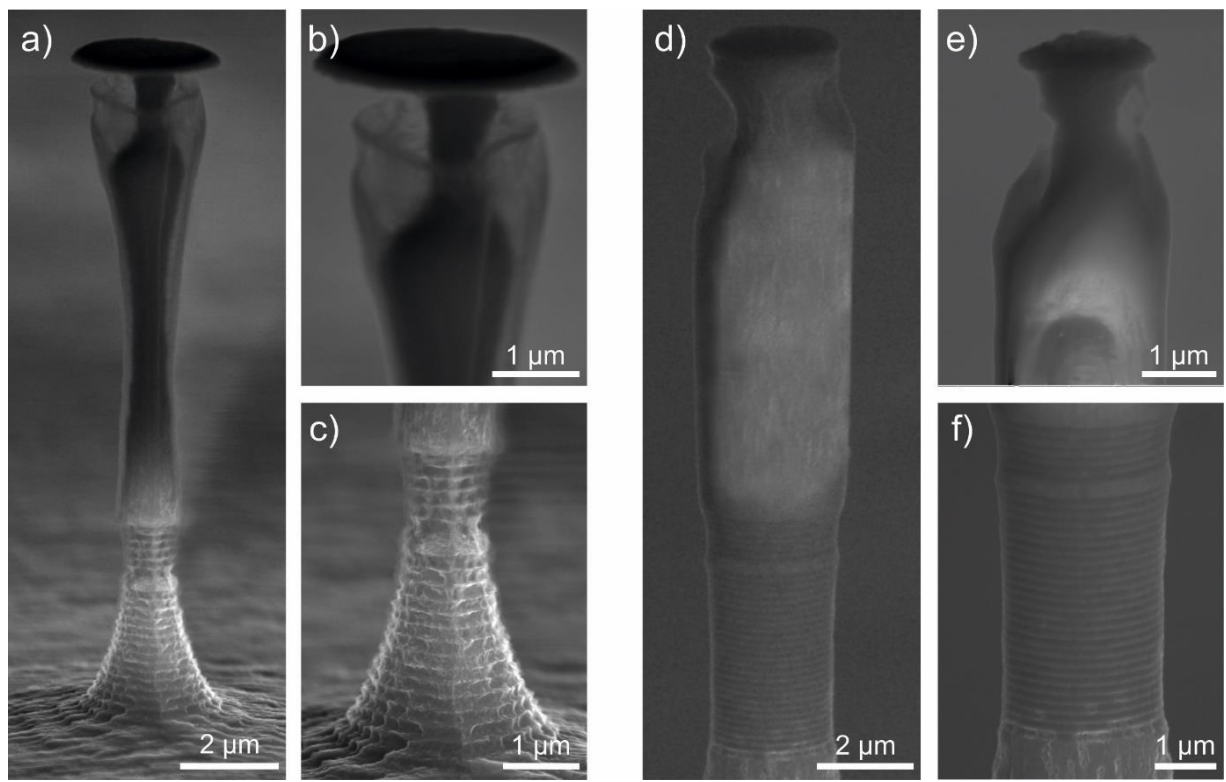

*Figure 3) SEM Images of processed structures: a) Etching the device on an undoped GaAs wafer yields an hourglass shape. Two unwanted process artefacts can be seen: first, a large under-etching beneath the Cr/BaF mask and second the oxidation and significant roughening on the sidewalls (both detailed in b). The oxidation is also visible in the DBR section of the structure (detailed in c). Here, also an unwanted pronounced etching of the $Al_{0.85}Ga_{0.15}As$ layers is clearly visible. d) Images of a structure etched on a doped wafer. The shape is rather straight and has a constant angle within the whole device. The under etching in the surrounding of the mask is less (detailed in e) than in the case of the undoped substrate. The DBR of this structure does not show selective etching and less sidewall oxidation (detailed in f) compared to the undoped wafer.*

To achieve the desired device shape, a systematic variation of the growth and the etching parameters was performed, as we describe in the following. The top diameters of the Bier-

Glass structures were defined by electron beam (e-beam) lithography with a diameter variation in the range of 0.8 µm to 2.4 µm. Afterwards, an etching mask of Cr/BaF was evaporated on the surface. In the next step, the remaining photoresist was removed with a lift-off process and the pattern was transferred to the planar sample by using reactive electron cyclotron resonance (ECR) plasma etching. Since the plasma etch process sensibly reacts on charge transport conditions in the sample, we first check the shape of nominally identical structures grown on a Si doped GaAs wafers as well as on undoped GaAs wafers with a SEM, shown in Figure 3. Figure 3a-c) shows a 14 µm high device, based on the heterostructure grown on an undoped GaAs wafer. The angle of the upper section (approx. 3°) is approximately ten times smaller than the angle in the lower DBR (approx. 25°) and increases towards the top of the taper section. The whole structure is covered with a transparent oxidation layer as also seen at micropillar structures before [24]. The thickness of this oxidation layer varies between 150 nm in the GaAs taper section and up to 280 nm in the DBR section. While the exact composition of the coating is not fully clear, it has been attributed to a re-deposition effect stemming from remains of the $SiO_2$ sample holder utilized in the etching chamber [24], similar to the $SiO_2$ mask used during the etching process of waveguides [25]. The structure is subject to a significant under-etching beneath the etching mask (Figure 3b)). The etching mask is visible as a black disk on top of the device (3 µm diameter). The remaining part of the GaAs right under the mask is approx. 850 nm thick and increases over a length of approx. 2 µm to a diameter of 1.5 µm. The behavior of the etching in the DBR segment also displays some peculiarities (Figure 3c): The DBR also shows strong selective oxidation and a rather rough surface with a pronounced modulation as seen before in etching processes[26]. We believe this modulation stems from a different etching speed of the AlGaAs and GaAs layers. The etching as well as the oxidation of the $Al_{0.85}Ga_{0.15}As$ layers is more pronounced due to the chemical etching and reactions of the residual etching products at the surface. The etch-anisotropy yields steps between the AlGaAs and GaAs layers as large as 100 nm and a step between the upper DBR and the GaAs taper. The waist of the structure has a width of approx. 700 nm and is placed approx. 500 nm vertically above the GaAs cavity layer. We note, that among multiple devices fabricated with this process, the minimal obtainable waist size was 500 nm. Devices with smaller diameters systematically broke during the etching step due to mechanical stress, which most likely resulted from strong selective oxidation.

Utilizing the identical etching recipe on heterostructures grown on a n-doped (Si) wafer with a doping concentration between $1.7-3 \times 10^{18}$ $cm^{-3}$ yields completely different results (Figure 3d-f)). The structure shows a rather straight body with only a small angle of approx. 3°, which is constant from the top to the bottom of the device. Compared to the structure grown on an undoped substrate, the thickness of the coating on the sidewalls is significantly reduced (approx. 50 nm) and uniform over the whole device. The under etching at the region of the mask is still visible (Figure 3d-e)) but not as pronounced as in the undoped wafer case. Furthermore, the DBR section, detailed in Figure 3f), does not indicate selective etching of the AlGaAs and GaAs layers, except for a slightly highlighted cavity area.

Since these results show very undesired features on undoped substrates, further device optimization was performed on n-doped substrates only. We optimized the previous etching recipe to get a more uniform sidewall angle and smaller diameters in the lower area of the device, as well as to reduce the persisting under etching beneath the Cr/BaF etching mask. The optimization took place on an inductively coupled plasma (ICP) etching. We found the best results for an etching composition of 2.3 $Cl_2$, $A_2$, a high-frequency power of 50 W and an ICP power of 300 W. The devices are compiled in Figure 4. The devices on this sample have a variation in top diameter between 7.7 µm and 1 µm, allowing us to study the changes in the shape, as well as the differences in the optical properties of the devices. The device in Figure 4a) has a top diameter of 1.25 µm, a sidewall angle of 2° around the cavity region which straightens towards the top, and a height of approx. 15.9 µm. The shape of the device, indeed resembles the canonical German Bier-Glass. The top of the Bier-Glass (Figure 4b)) still features rough sidewalls to a depth of 4 µm under the etching mask. We believe that this results from charging of the device under the mask during the etching process, which causes a different etching result in this area. Another possibility involves the fact, that the intrinsic oxide deposition from the sample holder is less efficient on the top of the device and causes a less pronounced sidewall cover.

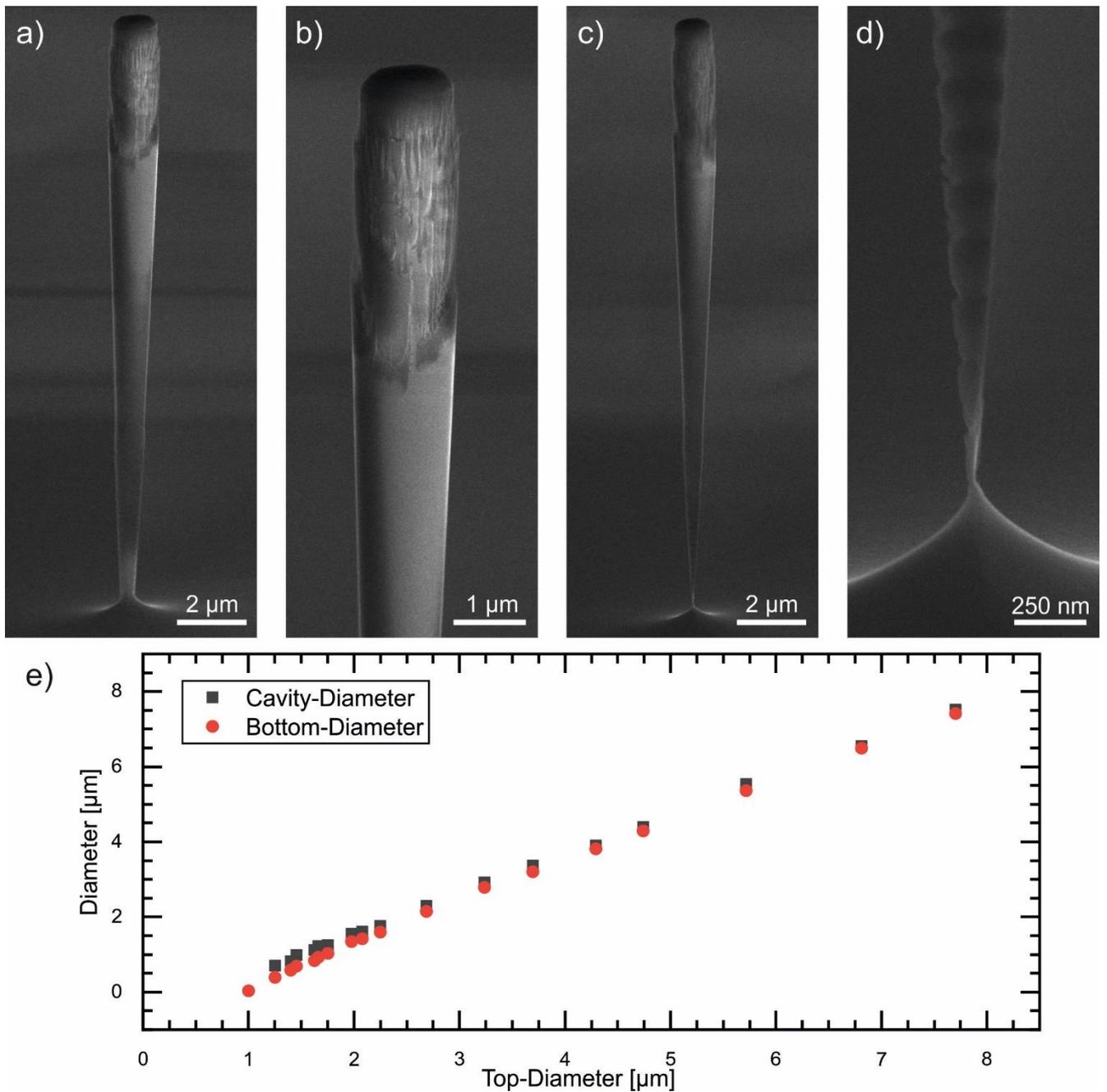

*Figure 4) SEM images of devices fabricated with optimized etching recipes. a) A Bier-Glass structure with a top diameter of 1.25 µm and a bottom diameter of approx. 400 nm. b) detail of the top from a) The top part of the device has a rougher sidewall in the vicinity of the Cr/BaF etching mask, which is the black cover on the pillar top surface. c) With this technique pillars with a top diameter of 1 µm and a diameter as low as approx. 20 nm at the thinnest part (detailed in d) can be achieved. e) The diameter at the position of the cavity layer as well as at the narrowest position of the Bier-Glass at the transition to the foot is plotted as a function of the top device diameter.*

As can be seen in Figure 4, the shape of the overall device remains identical for a reduced top diameter of 1 µm (Figure 4c)). Importantly, since the effects of etching selectivity between the GaAs and the AlGaAs mirrors and oxidation are strongly reduced with the optimized etching, the strain that builds up in the DBR segment of the device is dramatically reduced (compared to the devices which are presented in Figure 3a) and Figure 3b)). Therefore, the device stability is significantly improved, allowing us to fabricate Bier-Glasses with foot diameters as small as 20 nm. To the best of our knowledge, such an extreme height to base relation has not been achieved in any dry etched GaAs-based cavity structure before. All the device diameters (1 – 7.7 µm top diameter) have the same shape, but the visibility of the Bier-Glass shape is mostly pronounced for small top-diameters because of the better contrast between taper and foot. The device angle for all device is between 1.2° and 3°, which leads to a linear dependence (Figure 4e)) between top diameters and the narrowest diameter at the foot of the device. A small deviation from the linear dependence is visible for top diameters smaller than 2.5 µm.

The BaF/Cr hard mask is blocking light from the Bier-glass, hindering its use as a photon source. Thus, a crucial step to conduct optical experiments is to remove the BaF/Cr hard mask from the top. The hard mask is soluble in water, but this causes additional sidewall oxidation, especially in the Aluminum containing layers of the DBR sections, and significantly damages the devices. A process that protects the sidewall while washing off the etch mask was developed in previous studies [27]. This process involves spin coating of liquid benzocyclobutene (BCB), which is subsequently hardened by a baking process. Afterwards the mask can be washed away in water. Unfortunately, the structures with thinner bottom diameters break during the spin-coating, because of the critical aspect ratio of the Bier-Glass devices. To avoid spin-coating, we developed a technique that drops BCB onto the sample and lets it flow around the devices. To enhance the confinement of the liquid BCB, that would flow off the sample, protective walls were implemented (see Figure 5)). This technique allows us to protect the DBR part of the structure and wash away the hard mask. The disadvantage is that, the thickness of the BCB is not as uniform as with spin-coating. In Figure 5b) one can see planarized Bier-Glasses. The non-uniformity of the BCB is also visible notably in the vicinity of the pillars.

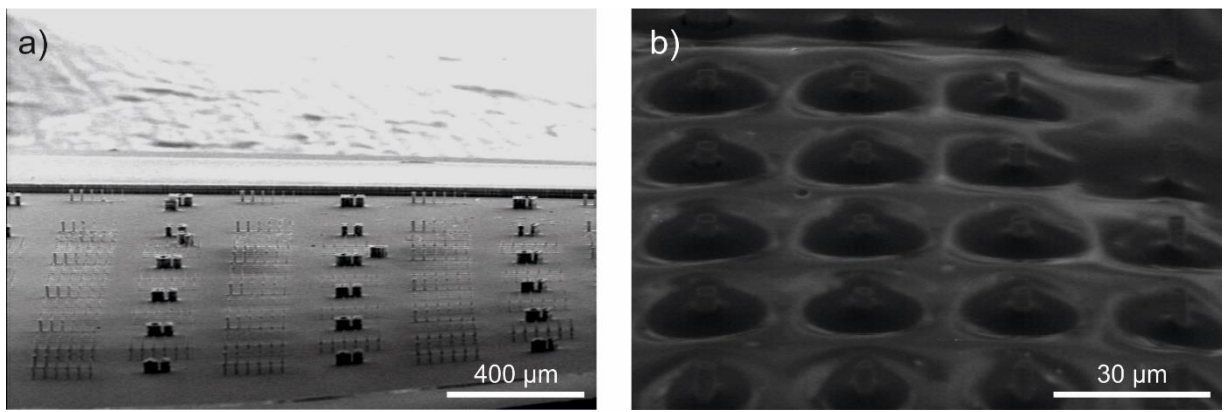

*Figure 5 a) global SEM image of the etched sample, including the protection wall to reduce the BCB flow. b) SEM image of the successfully planarized sample.*

**Optical characterization**

To characterize the optical performance of the devices, experiments in a micro-photoluminescence setup (µPL) were performed. For the excitation of the devices we used above band excitation utilizing a green continuous wave (CW) laser with a wavelength of 532 nm. The used objective has an NA of 0.42 and gives a focused spot size of 4.8 ± 0.7 µm. The sample was mounted in a helium flow cryostat and cooled down to about 10 K. In Figure 6a) we exemplarily show a µPL spectrum of a device with a large top diameter of 7.7 µm. High excitation power was used to saturate all the single quantum dot transitions and to get access to the device mode structure. At 750 µW excitation power we see that the lines start to redshift, which is an indication for local heating above 15 K. We find at least six modes visible in the range from 935 nm to 985 nm, with a mode spacing of 9 ± 1 nm. This mode spectrum is created by the Fabry-Pérot cavity formed between the surface to air and the upper DBR. To reduce the impact of these undesired features, we deposited a $Si_3N_4$ antireflection (AR) coating with a thickness of 126 nm on top of the sample [28]. The mode spectrum of the device measured in Figure 6a) after AR coating is shown in Figure 6b). It is clearly visible that the AR coating suppresses the Fabry-Pérot modes, enabling us to capture the fundamental mode of the 7.7 µm Bier-Glass device, with a central wavelength of 965 nm and a resonance linewidth of 17.4 ± 0.2 nm.

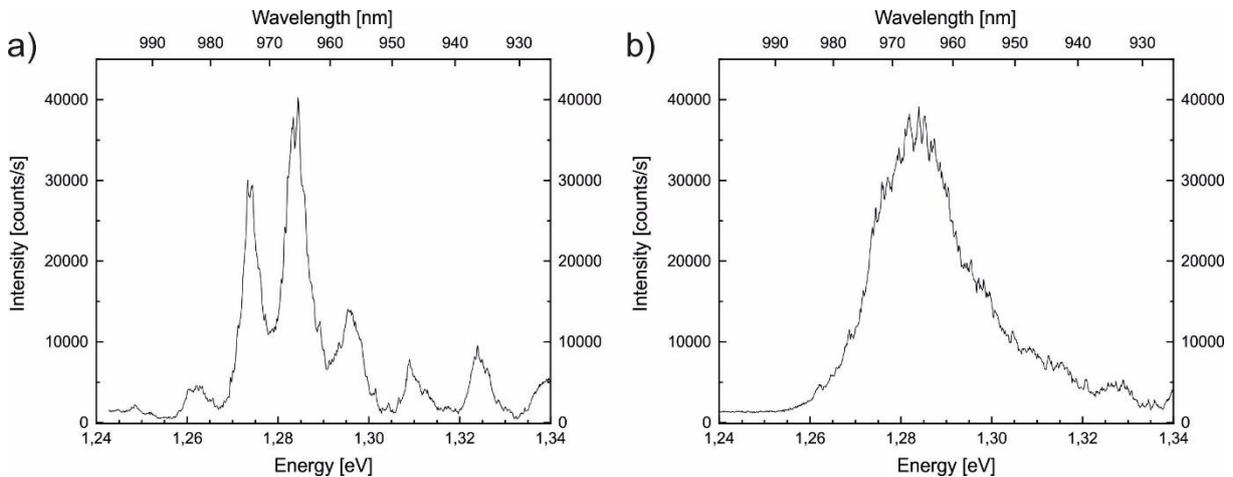

*Figure 6) Mode Spectra of a Bier Glass with a 7.7 µm top diameter. a) Spectra of the device without anti-reflective $Si_3N_4$ coating and b) with the anti-reflective coating. The fine spaced modes visible in a) are formed by a Fabry-Pérot resonator which is consisting of the upper DBR and the GaAs to air interface. b) $Si_3N_4$ deposition recovers the underlying mode of the Bier Glass device.*

Since the AR coating allows us to characterize the fundamental cavity mode, we study the dependence of the photonic confinement on its resonance energy (Figure 7a)). The fundamental cavity mode of the Bier-Glass device shifts to higher energies with reduced diameter, resulting from the lateral photonic confinement [29]. While the QD density in our structures was sufficiently large ($3 \times 10^9$ $cm^{-2}$) to homogenously illuminate the cavity resonances for large diameters, the cavity modes are strongly superimposed by single QD features in the smaller devices, adding some inaccuracy to determination of the mode energies, which are plotted as a function of the cavity diameter in Figure 7b). The error bars are the statistic variance of various results from different pillars with the same size.

While for the smallest devices, the experimental complications yield some increased fluctuations in the determined mode energies, the model, which is described above, nevertheless successfully reproduces the experimental shifts of the mode energies and supports our assignment of the broadband optical features to the fundamental cavity resonances.

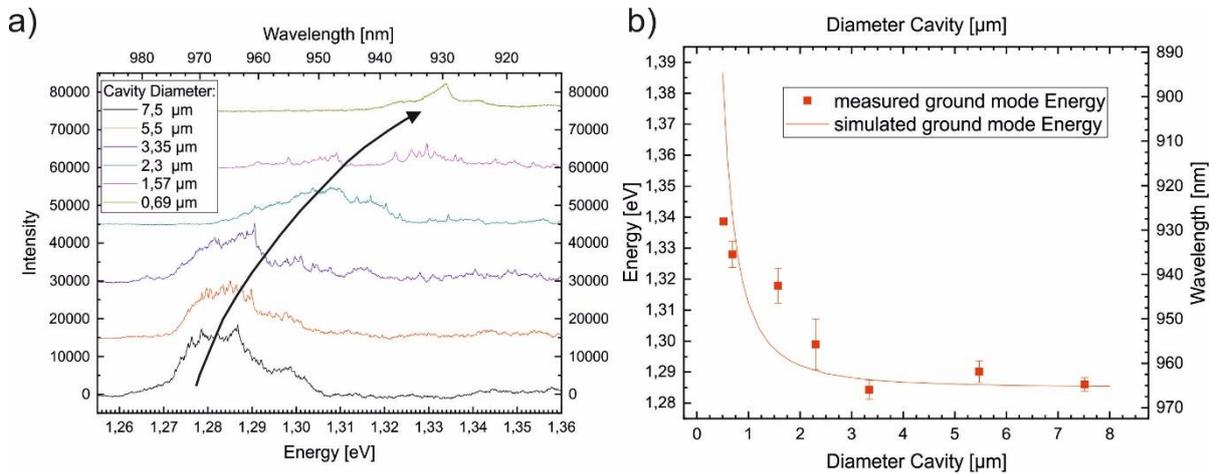

*Figure 7 a) Cavity Mode Spectra for different Bier-Glass top diameters. A clear shift to higher energies is visible and highlighted by a black arrow. b) Theoretical data for the ground mode energy (full orange line) and measured values for various diameters at the cavity waist.*

To assess the performance of our structure as a quantum optical device, we studied single QD transitions in the device. The power-dependent emission of QDs in a device with a top diameter of 1.6 µm can be seen in Figure 7a). We ramped the excitation power from 10 nW up to 30 µW where the QDs transitions are saturating. For further analysis, the two QDs indicated by the black arrows are investigated. The low energy QD at 943 nm is resonant with the fundamental cavity mode of the device, while the high energy QD (at 900 nm) is spectrally far detuned from the cavity resonance. In Figure 8b) the normalized integrated intensity for the two QDs is plotted as a function of the excitation power. The saturation behavior of both emitters can be approximated by the saturation of a two level system as [30,31] $I = I_{sat} (1/1+(P_n/P_{exc}))$ with the excitation power $P_{exc}$ and $P_n$ as a fitting parameter to normalize the excitation power. The fits are the solid lines in Figure 8b). The saturation power and the saturation intensity of the QD in resonance with the cavity mode is used to normalize the excitation power and the integrated intensity. The saturation intensity of the resonant QD is almost twenty times larger than the saturation intensity of the off-resonant QD. Since both QDs are within the same waveguide mode, this intensity enhancement can be attributed to the Purcell effect. The off-resonant QD is suppressed by the Bier-Glass structure, because in the waveguide the bulk mode density is strongly reduced. Thus, the Purcell factor cannot be extracted from the data, but we get an upper bound of 3 from theory. The comparison of the Bier-Glass device to a QD reference sample with QDs in bulk is also shown in Figure 8b). Here, the QD resonant to the Bier-Glass structure reaches a saturation intensity ten times higher than the emitter in the reference structure, which underlines the impact of the Bier-Glass structure on the photon extraction. The fact that the off resonant quantum dot is weaker than the reference structure, although it is emitting into a waveguide is underlining the suppression of the off resonant transition.

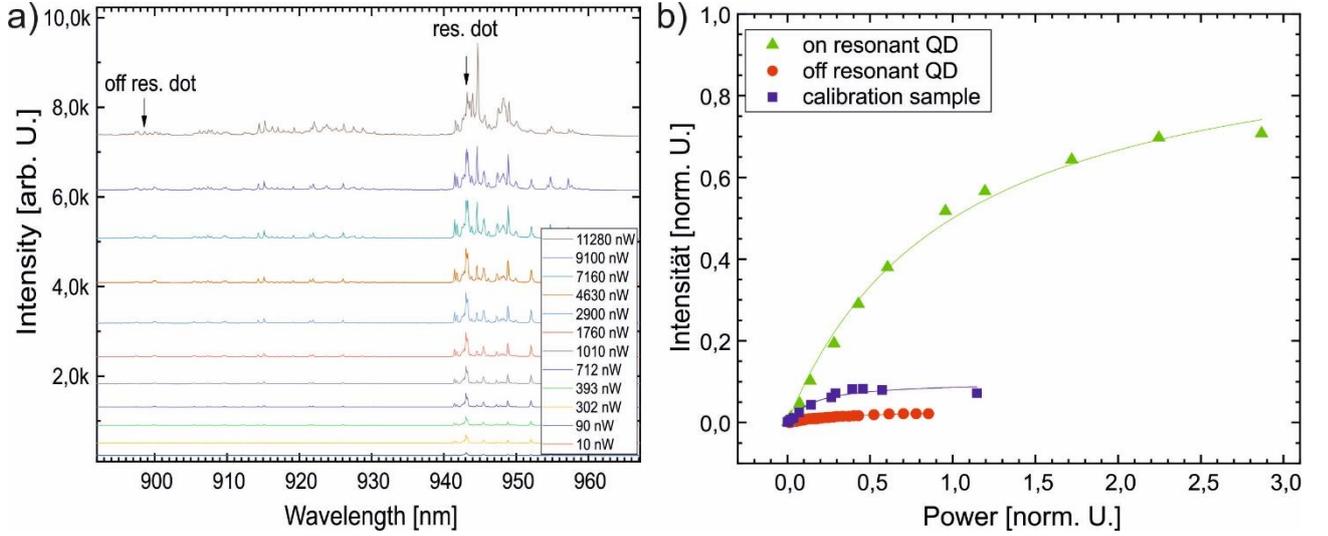

*Figure 8 a) The plot contains the intensity of a Bier-Glas Device with a top diameter of 1.6 µm for different excitation powers. Also indicated are two dot lines, one resonant and the other one off resonant. The Intensity (integrated under peak area) vs. the excitation power for those two lines is compared in Figure 8 b). The plots are fitted to extract the saturation power and intensity. The plot also contains a comparison of the Bier-Glas device with a planar QD calibration sample in the same way.*

Finally, we confirm the capability of our Bier-Glass cavities to act as non-classical light sources by measuring their second order auto-correlation function on a device with a top diameter of 3.6 µm. We excited the QD highlighted in Figure 9b), with a 532 nm CW laser at a power of 16 µW and passed it to an autocorrelation setup. The coincidences show the expected antibunching behavior at zero time delay (Figure 9a)). The fitting of the data with reveals a $g^{(2)}_{conv.}(0) = 0.366 \pm 0.10$ We assume that the value is non-zero, because of the finite time resolution of the detectors. To extract the correct value for g²(0), we convolve the expected g⁽²⁾ function with the measured system time response. The resulting fitting formula is [32]

$$g^{(2)}(\tau) = [1 - \left(1 - g^{(2)}_{deconv.}(0)\right) e^{\frac{-|\tau|}{\tau QP}}] \otimes G(t, tres)$$

and yields a value of $g^{(2)}_{deconv.}(0) = 0 ^{+0.11}_{-0.00}$. Both fitting functions are plotted together with the data in Figure 9 a).

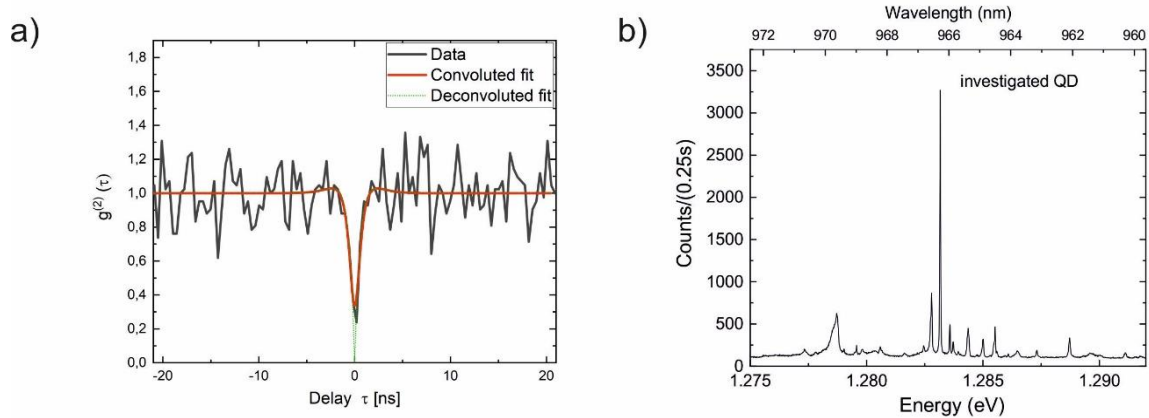

Figure 9 a) Coincidences vs. the time delay is fitted with the second order correlation function. The convoluted fit gives a value of $g^{(2)}_{conv.}(0) = 0.366 \pm 0.106$ for a time delay of 0 ns. After deconvolution of the fit the value for zero delay is $g^{(2)}_{deconv.}(0) = 0 \pm^{0.11}_{0.00}$. b) The full PL spectrum of the Bier-Glass selected for the $g^2(\tau)$ measurement.

## Conclusion

In conclusion, we have picked up an idea formulated by Gregersen et al. [13] to develop a device strategy for a single photon source that simultaneously provides broadband emitter-waveguide coupling with a broadband cavity resonance. Our systematic device optimization reflects the importance of the doping level of the substrate on the etching performance, the principal capability to fabricate Bier-Glass shaped structures featuring a desirable, long taper section of approx. 15 µm with a pedestal as thin as 20 nm, and we introduce a methodology to planarize such fragile objects with a polymer.

Our optical characterization confirms the presence of cavity modes as well as the improved coupling of single photons in our devices.

We are confident, that further adaption of the layer sequence and etching process will yield structures with ultra-large broadband efficiencies with substantial Purcell enhancement. Furthermore, our technological advancement, allowing to produce DBR-based cavities supported by 20 nm foots certainly can pave the way towards quantum-opto-mechanic applications.

## Acknowledgement

We gratefully acknowledge financial support by the state of Bavaria. The project HYPER-U-P-S has received funding from the QuantERA ERA-NET Co-fund in Quantum Technologies implemented within the European Union's Horizon 2020 Programme. We acknowledge Technical assistance by A. Wolf and M. Emmerling, as well as support by Dr. C. Anton-Solanas during the characterization of the devices.